\documentstyle[aps,prd,floats,epsf]{revtex}
\tighten
\draft
\begin{document}

\twocolumn[\hsize\textwidth\columnwidth\hsize\csname
@twocolumnfalse\endcsname
\title{The angular distribution of the reaction
$\bar{\nu}_e + p \rightarrow e^+ + n$}
\author{P. Vogel\thanks{Electronic address:
        {\tt vogel@lamppost.caltech.edu}} and
        J.~F. Beacom\thanks{Electronic address:
        {\tt beacom@citnp.caltech.edu}}}
\address{Physics Department 161-33, California
Institute of Technology, Pasadena, California 91125}
\date{March 31, 1999}
\maketitle

\begin{abstract} 

The reaction $\bar{\nu}_e + p \rightarrow e^+ + n$ is very important
for low-energy ($E_\nu \lesssim 60$ MeV) antineutrino experiments.  In
this paper we calculate the positron angular distribution, which at
low energies is slightly backward.  We show that weak magnetism and
recoil corrections have a large effect on the angular distribution,
making it isotropic at about 15 MeV and slightly forward at higher
energies.  We also show that the behavior of the cross section and the
angular distribution can be well-understood analytically for $E_\nu
\lesssim 60$ MeV by calculating to ${\cal O}(1/M)$, where $M$ is the
nucleon mass.  The correct angular distribution is useful for
separating $\bar{\nu}_e + p \rightarrow e^+ + n$ events from other
reactions and detector backgrounds, as well as for possible
localization of the source (e.g., a supernova) direction.  We comment
on how similar corrections appear for the lepton angular distributions
in the deuteron breakup reactions $\bar{\nu}_e + d \rightarrow e^+ + n
+ n$ and $\nu_e + d \rightarrow e^- + p + p$.  Finally, in the
reaction $\bar{\nu}_e + p \rightarrow e^+ + n$, the angular
distribution of the outgoing neutrons is strongly forward-peaked,
leading to a measurable separation in positron and neutron detection
points, also potentially useful for rejecting backgrounds or locating
the source direction.

\end{abstract}


\vspace{0.25cm}]
\narrowtext

%%%%%%%%%%%%%%%%%%%%%%%%%%%%%%%%%%%%%%%%%%%%%%%%%%%%%%%%%%%%%%%%%%%%%%%%%%%%
%%%%%%%%%%%%%%%%%%%%%%%%%%%%%%%%%%%%%%%%%%%%%%%%%%%%%%%%%%%%%%%%%%%%%%%%%%%%

\section{Introduction}

Inverse neutron beta decay, $\bar{\nu}_e + p \rightarrow e^+ + n$, is
the reaction of choice for the detection of reactor antineutrinos,
crucial to neutrino oscillation searches.  It is also, by far, the
reaction giving the largest yield for the detection of supernova
neutrinos.  The LSND~\cite{LSND} and KARMEN~\cite{KARMEN} experiments
use $\bar{\nu}_\mu$ antineutrinos from $\mu^+$ decay at rest to search
for the oscillation appearance of $\bar{\nu}_e$ events, detected by
this reaction.  There are also searches for $\bar{\nu}_e$
antineutrinos from the Sun.

In many of these applications, in particular those based on detection
by \v{C}erenkov radiation, one can determine the direction of the
outgoing positron.  It is therefore of interest to consider the
angular correlation between the incoming antineutrino and outgoing
positron directions, and its energy dependence for $E_\nu \lesssim 60$
MeV, relevant for the above studies.  If the source direction is
known, the angular correlation can be used to help separate these
events from other reactions or detector backgrounds.  If the source
direction is unknown, as possibly for a Galactic supernova, then the
observed angular distribution may help to locate the source.

For low antineutrino energies, the positron angular distribution
is well-described by the form
\begin{equation}
\frac{{\rm d}\sigma}{{\rm d }\cos\theta}
\simeq 1 + v_e a(E_\nu)\cos\theta\,,
\label{eq:angdist}
\end{equation}
where $\theta$ is the angle between the antineutrino and positron
directions in the laboratory (where the proton target is assumed to be
at rest) and $v_e$ is the positron velocity in $c = 1$ units.  At
higher energies, terms proportional to higher powers of $\cos\theta$
appear, and at the highest energies, all reaction products are
strongly forward simply by kinematics.  It is convenient to describe
the angular distribution by the average cosine, weighted by the
differential cross section, since that is always well-defined.  In the
limit that Eq.~(\ref{eq:angdist}) holds,
\begin{equation}
\langle \cos\theta \rangle = \frac{1}{3} v_e a(E_\nu)\,,
\end{equation}
where the $E_\nu$-dependence of $\langle \cos\theta \rangle$ is
suppressed in the notation here and below.  Except near threshold,
where $v_e$ becomes very small (but nonzero in the lab), the $v_e$
factor is nearly unity and can be ignored.

In the limit where the nucleon mass is taken to be infinite, i.e.,
zeroth order in $1/M$, the asymmetry coefficient $a$ is independent
of $E_\nu$ and would be the same for $\bar{\nu}_e + p \rightarrow e^+
+ n$ and $\nu_e + n \rightarrow e^- + p$.  (Since there are no free
neutron targets, the latter cannot be directly observed.  We return to
this point below in discussing neutrino and antineutrino reactions
with deuterons.)  Then $a^{(0)}$ is given simply by the competition of
the non-spin-flip (Fermi) and spin-flip (Gamow-Teller) contributions,
and is
\begin{equation}
a^{(0)} = \frac{f^2 - g^2}{f^2 + 3g^2} \simeq -0.10\,,
\label{eq:a0}
\end{equation}
and thus the angular distribution of the positrons is weakly backward.
We have defined the vector and axial-vector coupling constants by $f =
1$, $g = 1.26$.

In the following we will consider how $\langle \cos \theta \rangle$ is
modified when weak magnetism and recoil corrections of ${\cal O}(1/M)$
are kept.  The effect of these terms on the total cross section was
calculated in Refs.~\cite{Vogel,Fayans1} (see also
Ref.~\cite{Seckel}), where it was found that they, in particular the
weak magnetism, reduce the cross section by a noticeable amount.  In
this paper, we will show that the positron angular distribution is
changed even more, including the sign of $\langle \cos \theta
\rangle$.  The effect is so large in part because of the accidental
near-cancelation in $a^{(0)}$.

The general form of the differential cross section, valid to all
orders in $1/M$ but neglecting the threshold effects (and hence only
valid for energies far above threshold), is well-known~\cite{LS}.  For
the relevant energies $E_\nu \lesssim 60$ MeV, it is instructive and
sufficient to consider in detail just the terms of first order in
$1/M$.  Here and below, $1/M$ will be taken to refer to all terms of
that form, with $E_\nu/M$ being dominant among them.  Moreover, using
these results, we show how to extend the formula of Ref.~\cite{LS} to
low energies, so that it merges smoothly with the correct expression
near threshold.

%%%%%%%%%%%%%%%%%%%%%%%%%%%%%%%%%%%%%%%%%%%%%%%%%%%%%%%%%%%%%%%%%%%%%%%%%%%%
%%%%%%%%%%%%%%%%%%%%%%%%%%%%%%%%%%%%%%%%%%%%%%%%%%%%%%%%%%%%%%%%%%%%%%%%%%%%

\section{The positron angular distribution}

%%%%%%%%%%%%%%%%%%%%%%%%%%%%%%%%%%%%%%%%%%%%%%%%%%%%%%%%%%%%%%%%%%%%%%%%%%%%

\subsection{Differential cross section: expansion in powers of $1/M$}

We begin with the matrix element of the form
\begin{eqnarray}
{\cal M} & = &
\frac{G_F \cos \theta_C}{\sqrt{2}}
\left[ \bar{u}_n \left(\gamma_{\mu} f - \gamma_{\mu}\gamma_5 g -
\frac{i f_2}{2 M} \sigma_{\mu \nu} q^{\nu} \right) u_p \right] \nonumber \\
& \times &
\left[ \bar{v}_{\bar{\nu}} \gamma^{\mu} ( 1 - \gamma_5 ) v_e \right]\,,
\label{eq:amp}
\end{eqnarray}
where $f$ and $g$ are given above, the anomalous nucleon isovector
magnetic moment is defined with $f_2 = \mu_p - \mu_n = 3.706$, and
$\cos\theta_C = 0.974$.  In the most general case, the coupling
constants are replaced with form factors that vary with $q^2$; we
neglect this variation as it is ${\cal O}(E_{\nu}^2/M^2)$.  The
four-momentum transfer $q^2$ is related to the laboratory scattering
angle $\theta$, which in turn is related to the outgoing positron
energy $E_e$ (again in the laboratory) by the relations
\begin{eqnarray}
q^2 & = & m_e^2 - 2 E_{\nu} E_e ( 1 - v_e \cos \theta ) \\
& = & (M_n^2 - M_p^2) - 2 M_p ( E_{\nu} - E_e )\,.
\end{eqnarray}
Some other useful kinematic relations are given in the Appendix.  We can
now use the standard rules and evaluate the differential cross section
accurate to a given order in $1/M$.

At zeroth order in $1/M$, the positron energy is
\begin{equation}
E_e^{(0)} = E_\nu - \Delta,
\end{equation}
where $\Delta = M_n - M_p$.  At each order in $1/M$, we define the
positron momentum $p_e = \sqrt{E_e^2 - m_e^2}$ and the velocity $v_e =
p_e/E_e$.  The differential cross section at this order is
\begin{eqnarray}
& & \left(\frac{{\rm d}\sigma}{{\rm d cos}\theta}\right)^{(0)}
\nonumber \\
& = &
\frac{\sigma_0}{2}
\left[(f^2 + 3g^2) + (f^2 - g^2) v_e^{(0)} \cos\theta \right]
E_e^{(0)} p_e^{(0)}\,.
\label{eq:dsig0}
\end{eqnarray}
The normalizing constant $\sigma_0$, including the energy-independent
inner radiative corrections, is
\begin{equation}
\sigma_0 = \frac{G_F^2 \cos^2\theta_C}{\pi}\; (1 + \Delta^R_{inner})\,,
\end{equation}
where $\Delta^R_{inner} \simeq 0.024$~\cite{RC}.  This gives the
standard result for the total cross section,
\begin{eqnarray}
\sigma^{(0)}_{tot} & = &
\sigma_0\; (f^2 + 3 g^2)\; E_e^{(0)} p_e^{(0)} \\
& = & 0.0952 \left(\frac{E_e^{(0)} p_e^{(0)}}{1 {\rm\ MeV}^2}\right)
\times 10^{-42} {\rm\ cm}^2\,.
\label{eq:sigtot0}
\end{eqnarray}

The energy-independent inner radiative corrections affect the neutron
beta decay rate in the same way, and hence the total cross section can
also be written
\begin{equation}
\sigma^{(0)}_{tot} = 
\frac{2 \pi^2/m_e^5}{f^R_{p.s.} \tau_n}\; E_e^{(0)} p_e^{(0)}\,,
\end{equation}
where $\tau_n$ is the measured neutron lifetime and $f^R_{p.s.} =
1.7152$ is the phase space factor, including the Coulomb, weak
magnetism, recoil, and outer radiative corrections, but {\it not} the
inner radiative corrections~\cite{Wilkinson}.  The cross section
normalization was measured in Ref.~\cite{Declais} and found to be in
agreement with the expectation from the neutron lifetime at the 3\%
level.  The (small) energy-dependent outer radiative corrections to
$\sigma_{tot}$ are given in Refs.~\cite{Vogel,Fayans1}.  The outer
radiative corrections to $\langle \cos\theta \rangle$ should largely
cancel in the ratio of the cross section weighted with $\cos\theta$ to
the cross section itself, and so are not considered further here.

At first order in $1/M$, the positron energy depends upon the
scattering angle and is
\begin{equation}
E_e^{(1)} =
E_e^{(0)}
\left[1 - \frac{E_{\nu}}{M}(1 - v_e^{(0)} \cos \theta) \right]
- \frac{y^2}{M}\,,
\label{eq:eel1}
\end{equation}
where $y^2 = (\Delta^2 - m_e^2 )/2$. In factors of the form $1/M$, we
use the average nucleon mass; using $1/M$ versus $1/M_p$ leads to an
ignorable difference of ${\cal O}(1/M^2)$.  The differential cross
section at this order (after a lot of tedious algebra) is
\begin{eqnarray}
& & \left(\frac{{\rm d}\sigma}{{\rm d cos}\theta}\right)^{(1)}
\nonumber \\
& = & \frac{\sigma_0}{2}
\left[\vphantom{\frac{\sigma_0}{2}}
(f^2 + 3g^2) + (f^2 - g^2) v_e^{(1)} \cos\theta\right]
E_e^{(1)} p_e^{(1)} \nonumber \\
& - & \frac{\sigma_0}{2}
\left[\frac{\Gamma}{M}\right]
E_e^{(0)} p_e^{(0)}\,,
\label{eq:dsig1}
\end{eqnarray}
where
\begin{eqnarray}
\Gamma & = &
2 (f + f_2) g
\left[ (2 E_e^{(0)} + \Delta ) ( 1 - v_e^{(0)}\cos\theta)
-\frac{m_e^2}{E_e^{(0)}} \right] \nonumber\\
& + &
(f^2 + g^2)
\left[ \Delta( 1 + v_e^{(0)}\cos\theta)
+ \frac{m_e^2}{E_e^{(0)}} \right] \nonumber\\
& + & 
(f^2 + 3g^2)
\left[ (E_e^{(0)} + \Delta) (1 - \frac{1}{v_e^{(0)}}\cos\theta) - 
\Delta\right] \nonumber\\
& + &
(f^2 - g^2)
\left[ (E_e^{(0)} + \Delta) (1 - \frac{1}{v_e^{(0)}}\cos\theta) - 
\Delta\right] v_e^{(0)} \cos\theta \,.
\end{eqnarray}
For the dominant term (the first square brackets in
Eq.~(\ref{eq:dsig1})), the cosine-dependence of $E_e$, $p_e$, and
$v_e$ must be taken into account at ${\cal O}(1/M)$, while for the
subdominant terms, they may be taken as functions of $E_\nu$ alone.
Note that the terms $1/v_e^{(0)}$ above, which can be large near
threshold, are canceled by $p_e^{(0)}$ in the phase space factor.  Our
result for the total cross section, calculated from
Eq.~(\ref{eq:dsig1}), supercedes the result of Ref.~\cite{Vogel},
which did not include the ${\cal O}(1/M)$ corrections to the Jacobian
$dt/d\cos\theta$.

Unless the electron mass is negligible (see below), it is not
convenient to analytically expand $E_e^{(1)}$, $p_e^{(1)}$, and
$v_e^{(1)}$ in powers of $1/M$, and instead we evaluate the total
cross section $\sigma_{tot}^{(1)}$ numerically.  In the upper panels
of Figs.~1 and 2 we show the total cross section versus $E_\nu$.  We
divided the plots into two energy regimes: Fig.~1 to show the
threshold region, relevant for, e.g., reactor experiments; and Fig.~2
to show the global behavior relevant for, e.g., supernova or muon
decay at rest experiments.  The solid line is the result at first
order in $1/M$, given by numerical integration of
Eq.~(\ref{eq:dsig1}).  The short-dashed line is the result at zeroth
order in $1/M$, given by Eq.~(\ref{eq:sigtot0}).  As expected, these
results agree at the very lowest energies.  However, with increasing
energy the $1/M$ corrections become more and more important, reducing
the total cross section.

\subsection{Differential cross section: the high-energy limit}

Far above threshold, our result can be compared to Eq.~(3.18) of
Ref.~\cite{LS}; as noted, that formula neglects $\Delta$ but contains
all orders in $1/M$.  At low energies, the neglect of the threshold is
a large effect, as shown by the long-dashed line in the upper panels
of Figs.~1 and 2.  We have modified Eq.~(3.18) of Ref.~\cite{LS} to
take into account the largest contributions of the threshold effects.
First, the exact kinematics (see our Appendix), including $\Delta$,
should be used to evaluate $q^2$ and $s - u$ in that formula.  With no
further modification, that formula does not have the correct low
energy limit (determined by comparing to the results above).  By
direct comparison, the only other $\Delta$-dependent correction to the
formula of Ref.~\cite{LS} at ${\cal O}(1)$ is the modification
\begin{equation}
C(q^2) (s - u)^2 \rightarrow 
C(q^2) (s - u)^2 - C(q^2)\; 4 M^2 \Delta^2\,.
\end{equation}
As shown by the dot-dashed line in the upper panels of Figs.~1 and 2,
this corrects the result of Ref.~\cite{LS} so that it takes into
account the threshold.  There may be additional corrections of order
$\Delta/M$ necessary, but by the numerical results, they are evidently
small.

We took the form factors in Eq.~(\ref{eq:amp}) to be constants, which
is reasonable for the energies considered.  At higher energies, the
form-factor variation with momentum transfer must be properly
included, as done in \cite{LS}.  Note that the form-factor variation
in Ref.~\cite{Fayans1} is incorrect, as it attributes a dipole
behavior to the coefficients which appear directly in
Eq.(\ref{eq:amp}), and not to the momentum-transfer dependent linear
combinations of them known as the Sachs form factors; see
Ref.~\cite{LS}.

\begin{figure}[t]
\epsfxsize=3.25in \epsfbox{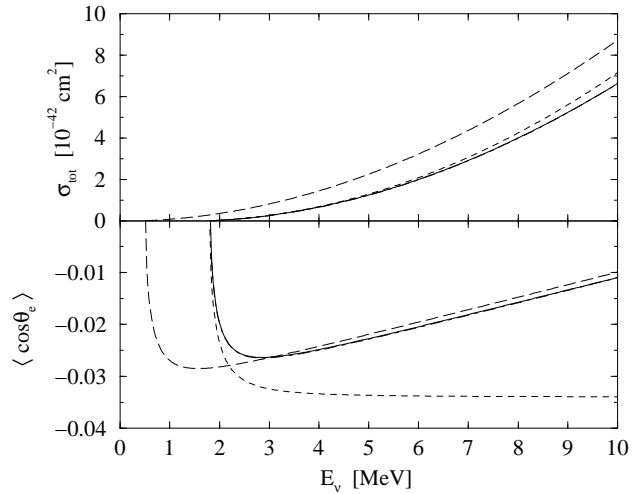}
\caption{Upper panel: total cross section for $\bar{\nu}_e + p
\rightarrow e^+ + n$; bottom panel: $\langle \cos\theta \rangle$ for
the same reaction; both as a function of the antineutrino energy.  The
solid line is our ${\cal O}(1/M)$ result and the short-dashed line is
the ${\cal O}(1)$ result.  The long-dashed line is the result of
Eq.(3.18) of Ref.~\protect\cite{LS}, and the dot-dashed line contains our
threshold modifications to the same.  The solid and dot-dashed lines
are not distinguishable in this figure.  The inner radiative
corrections are included (see the text), but the outer radiative
corrections are not (see Refs.~\protect\cite{Vogel,Fayans1}).}
\end{figure}

The plotted results show that our $\sigma_{tot}^{(1)}$ agrees well
with the modified results of Ref.~\cite{LS}, particularly at low
energies.  While lowest order in $1/M$ is clearly not enough, the
first order in $1/M$ is, justifying our neglect of higher orders.  One
can see that the short-dashed line differs from the others
substantially already at $E_{\nu} \gtrsim$ 30 MeV. This suggests that
in the expansion in $1/M$ the numerical coefficient multiplying the
dominant term $E_{\nu}/M$ is quite large ($\simeq - 7$).  One can see
that explicitly by examining Eq.~(\ref{eq:sigtot1m0}) below.

%%%%%%%%%%%%%%%%%%%%%%%%%%%%%%%%%%%%%%%%%%%%%%%%%%%%%%%%%%%%%%%%%%%%%%%%%%%%

\subsection{Angular distribution}

At zeroth order, recall that
\begin{equation}
\langle \cos\theta \rangle^{(0)} = v_e^{(0)} a^{(0)}/3
\simeq -0.034 v_e^{(0)}\,,
\label{eq:avgcos0}
\end{equation}
so the angular distribution is slightly backward, independent of
energy (above the threshold region).  From Eq.~(\ref{eq:dsig1}), it is
evident that the angular distribution will be modified by the $1/M$
corrections.  However, the corrections in $E_e^{(1)}$, $p_e^{(1)}$,
and $v_e^{(1)}$ are not explicitly shown.  At lowest order, the
$m_e$-dependent effects appear always as $m_e^2/E_e^2$, and for $E_\nu
\gtrsim 5$ MeV, these may be neglected so that
\begin{eqnarray}
& & \left(\frac{{\rm d}\sigma}{{\rm d cos}\theta}\right)^{(1)}
\nonumber \\
& \simeq &
\frac{\sigma_0}{2}
\left[\vphantom{\frac{\sigma_0}{2}}
(f^2 + 3g^2) + (f^2 - g^2) \cos\theta - \frac{\Gamma}{M}\right]
E_e^{(0)} E_e^{(0)}
\label{eq:dsig1m0}
\end{eqnarray}
where
\begin{eqnarray}
\Gamma & = &
2 (f + f_2) g
\left[ (2 E_e^{(0)} + \Delta ) ( 1 - \cos\theta)\right] \nonumber\\
& + &
(f^2 + g^2)
\left[ \Delta( 1 + \cos\theta)\right] \nonumber\\
& + & 
(f^2 + 3g^2)
\left[3 (E_e^{(0)} + \Delta) (1 - \cos\theta) 
- \Delta\right] \nonumber\\
& + &
(f^2 - g^2)
\left[3 (E_e^{(0)} + \Delta) (1 - \cos\theta) 
- \Delta\right] \cos\theta \,,
\end{eqnarray}
where in the latter two square brackets, we have also neglected terms
$+\Delta^2/E_e^{(0)}$.  Note that the $(f^2 + 3 g^2)$ and $(f^2 -
g^2)$ terms are modified from Eq.~(\ref{eq:dsig1}) by terms from the
expansion of the phase space factor $E_e^{(1)} p_e^{(1)}$.

It is now trivial to integrate over $\cos\theta$ and to determine
$\sigma_{tot}$ and the integral weighted with $\cos \theta$, which we
call $(d \sigma \cos\theta)_{tot}$.  These can be written as
\begin{equation}
\sigma_{tot}^{(1)} = 
\sigma_0 \left(\alpha_1 + \beta_1 \frac{\Delta}{M}
+ \gamma_1 \frac{E_e^{(0)}}{M}\right)
E_e^{(0)} E_e^{(0)}
\label{eq:sigtot1m0}
\end{equation}
and
\begin{equation}
(d \sigma \cos\theta)_{tot}^{(1)} = 
\sigma_0 \left(\alpha_2 + \beta_2 \frac{\Delta}{M}
+ \gamma_2 \frac{E_e^{(0)}}{M}\right)
E_e^{(0)} E_e^{(0)}\,.
\label{eq:costot1m0}
\end{equation}
The coefficients $\alpha,\beta,\gamma$ can be immediately read off of
Eq.~(\ref{eq:dsig1m0}), since the $\cos\theta$ integration is trivial.
In order to continue working consistently to order $1/M$, we divide
$(d\sigma \cos\theta)_{tot}/\sigma_{tot}$ analytically.  Since
$\gamma_1/\alpha_1 \simeq - 7$ is large, numerical division would
improperly introduce higher-order terms.  Then
\begin{eqnarray}
& & \langle \cos\theta \rangle^{(1)} \nonumber \\
& \simeq &
\frac{\alpha_2}{\alpha_1} 
\left[v_e^{(0)}
+  \left(\frac{\beta_2}{\alpha_2} - \frac{\beta_1}{\alpha_1}\right)
\frac{\Delta}{M}
+ \left(\frac{\gamma_2}{\alpha_2} - \frac{\gamma_1}{\alpha_1}\right)
\frac{E_e^{(0)}}{M}\right]\,.
\label{eq:avgcos1}
\end{eqnarray}
It can be shown both analytically and numerically that by far the
largest $m_e$-dependent effect can be restored by the insertion of the
term $v_e^{(0)}$ as above, since the phase-space factors $E_e p_e$
cancel in the definition of $\langle \cos\theta \rangle$.  This
formula is an excellent approximation for $\langle \cos\theta \rangle$
from threshold even to $E_\nu \simeq 150$ MeV (though at that energy
neither of Eqs.~(\ref{eq:sigtot1m0}) and (\ref{eq:costot1m0}) is
individually valid, and the angular distribution is no longer of the
form $1 + a \cos\theta$).

Keeping only the largest terms, we can also write
\begin{eqnarray}
\langle \cos\theta \rangle^{(1)} & \simeq &
\frac{v_e^{(0)} a^{(0)}}{3}
+ \frac{1}{3}
\left(3 + \frac{4 (f + f_2) g}{(f^2 + 3 g^2)}\right)
\frac{E_\nu}{M} \nonumber \\
& \simeq & -0.034 v_e^{(0)} + 2.4 \frac{E_\nu}{M}\,.
\label{eq:avgcos1a}
\end{eqnarray}
The standard $M \rightarrow \infty$ result is very small, $\langle
\cos\theta \rangle^{(0)} \simeq -0.034 v_e^{(0)}$.  The large
corrections depending on $E_\nu/M$ can be classified as being due to
weak magnetism (depending on $(f + f_2)$) and pure recoil (independent
of the couplings).  For $\bar{\nu}_e + p \rightarrow e^+ + n$, these
add.  For the reaction $\nu_e + n \rightarrow e^- + p$, the sign of
$(f + f_2)$ is reversed, and the recoil and weak magnetism terms would
nearly cancel.

\begin{figure}[t]
\epsfxsize=3.25in \epsfbox{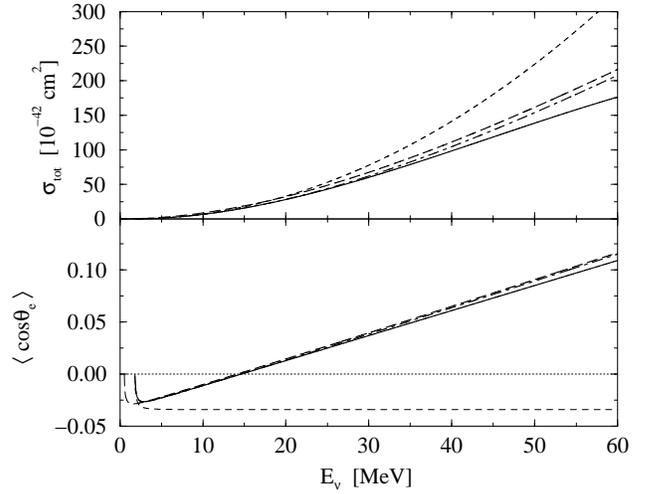}
\caption{The same as Fig.~1, but over a larger range of antineutrino
energy.  The long-dashed and dot-dashed lines are nearly
indistinguishable in the lower panel.}
\end{figure}

Our results for $\langle \cos\theta \rangle$ are shown in the lower
panels of Figs.~1 and 2.  Our main result at order $1/M$, given by
Eq.~(\ref{eq:avgcos1}), is shown as the solid line.  The zeroth order
result is shown as the short-dashed line.  The long-dashed line shows
the result of Eq.~(3.18) of Ref.~\cite{LS}, which assumes $\Delta =
0$.  This is obviously poor in the threshold region.  The dot-dashed
line shows our modification of that formula to account for the largest
$\Delta$-dependent effects.  Note that the upward ``hook'' at low
energies ($E_{\nu} \le 5$ MeV) is caused by the finite electron mass.
At those energies, $v_e < 1$, and the average $\langle \cos \theta
\rangle$ decreases, nearly vanishing as the antineutrino energy
approaches its threshold value.

These results for $\langle \cos\theta \rangle$ agree qualitatively
with the earlier numerical results of Ref.~\cite{Perkins}, which noted
that the $\langle \cos \theta \rangle$ vanishes near $E_{\nu} = 20$
MeV, and becomes slightly positive at larger antineutrino energies.
At lower energies, the results of Ref.~\cite{Perkins} are inaccurate,
presumably due to using the formula of Ref.~\cite{LS} without the
threshold modifications given above.

Terms of the first order in $1/M$ radically change $\langle \cos
\theta \rangle$, including its sign.  At high energies, the missing
$1/M^2$ terms become important for $\sigma_{tot}$ and (not shown) $(d
\sigma \cos\theta)_{tot}$.  However, note that $1/M^2$ effects are
negligible for $\langle \cos\theta \rangle$.  Our result, given by the
solid line in the lower panels of Figs.~1 and 2, contains {\it no}
terms of order $1/M^2$ or higher.  The modified result (with our
corrections for $\Delta > 0$) of Ref.~\cite{LS}, given by the
dot-dashed line, contains {\it all} terms of order $1/M^2$ and higher.
The agreement is excellent, and both are approximately linear in
$E_\nu/M$.  That is, for $\langle \cos\theta \rangle$, there is a
large cancelation of the higher order corrections.

%%%%%%%%%%%%%%%%%%%%%%%%%%%%%%%%%%%%%%%%%%%%%%%%%%%%%%%%%%%%%%%%%%%%%%%%%%%%

\subsection{Charged-current deuteron breakup reactions}

Since there are no free neutron targets, the reaction $\nu_e + n
\rightarrow e^- + p$ cannot be observed directly.  However, since the
deuteron is so weakly bound, we can at least qualitatively apply the
weak magnetism and recoil effects calculated above to the reactions
$\bar{\nu}_e + d \rightarrow e^+ + n + n$ and $\nu_e + d \rightarrow
e^- + p + p$.  For the considered energies, these reactions are pure
Gamow-Teller transitions, and so the asymmetry is $a^{(0)} = -1/3$.
In both reactions, $\langle \cos\theta \rangle$ will be made more
positive by pure recoil corrections.  To those, the weak magnetism
correction adds for $\bar{\nu}_e + d$ and subtracts for $\nu_e + d$.

In Fig.~3 we show the $\langle \cos \theta \rangle$ results calculated
from the double-differential cross sections of
Kubodera~\cite{pcKubodera}.  His results are based on a complete
calculation, including treatment of the deuteron wave function and
meson-exchange effects~\cite{Kubodera}.  The corrections due to the
finite nucleon mass are evident.  One can see that the two curves are
not symmetric with respect to the $M \rightarrow \infty$ value
$\langle \cos \theta \rangle = -1/9$.  As expected, the weak magnetism
and recoil corrections act in the same sense for $\bar{\nu}_e + d$ and
the opposite sense for $\nu_e + d$.

The weak magnetism contribution can be analytically estimated from the
amplitude squared in Ref.~\cite{Fayans2}.  We estimate the recoil
contribution so that the combined result
\begin{equation}
\langle \cos\theta \rangle^{(1)} \simeq
-\frac{1}{9}
+ \frac{1}{3} \left(2
\pm \frac{8 f_2}{9 g}\right) \frac{E_\nu}{M}
\end{equation}
provides a reasonable fit to the full numerical results in Fig.~3.

\begin{figure}[t]
\epsfxsize=3.25in \epsfbox{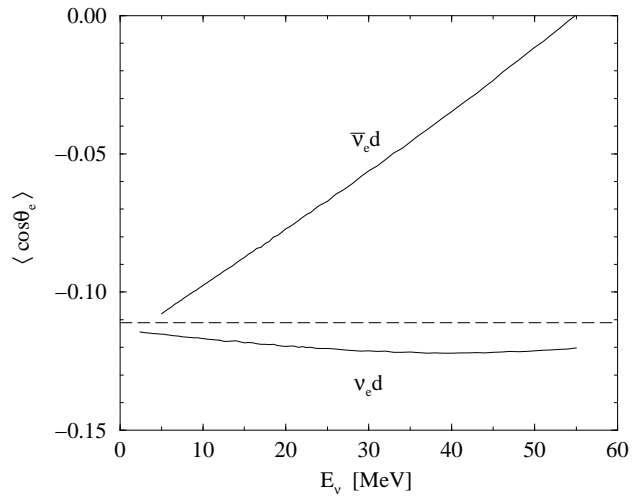}
\caption{The average lepton cosine for the charged-current deuteron
reactions, versus the neutrino or antineutrino energy, using
Kubodera's calculations.  We plot only from 1 MeV above threshold, so
the $v_e$-dependence at low energies is not shown.  The small jitters
are due to the coarse integration grid.}
\end{figure}

%%%%%%%%%%%%%%%%%%%%%%%%%%%%%%%%%%%%%%%%%%%%%%%%%%%%%%%%%%%%%%%%%%%%%%%%%%%%
%%%%%%%%%%%%%%%%%%%%%%%%%%%%%%%%%%%%%%%%%%%%%%%%%%%%%%%%%%%%%%%%%%%%%%%%%%%%

\section{Applications of the positron angular distribution}

%%%%%%%%%%%%%%%%%%%%%%%%%%%%%%%%%%%%%%%%%%%%%%%%%%%%%%%%%%%%%%%%%%%%%%%%%%%%

\subsection{SN1987A events}

Supernova 1987A was observed in two water-\v{C}erenkov detectors,
Kamiokande II~\cite{KamII} and IMB~\cite{IMB}, with 12 and 8 events,
respectively.  These events were presumably entirely due to
$\bar{\nu}_e + p \rightarrow e^+ + n$, with an angular distribution of
the form $1 + a \cos\theta$.  A well-known peculiarity of the SN1987A
data is that the angular distributions of the detected positrons are
apparently too forward, with $\langle \cos\theta \rangle = 0.34$ in
Kamiokande II and $\langle \cos\theta \rangle = 0.48$ in IMB.  Using
the results of Fig.~2, evaluated at the observed average energies, and
the correction for the IMB angular bias~\cite{IMB}, we would expect
only $\langle \cos\theta \rangle \simeq 0$ in Kamiokande II and
$\langle \cos\theta \rangle \simeq 0.08$ in IMB.

The error on the mean $\langle \cos\theta \rangle$ is
\begin{equation}
\delta\left(\langle \cos\theta \rangle\right)
= \frac{1}{\sqrt{N}} \sqrt{\frac{1}{3} - \frac{a^2}{3^2}}
\simeq  \frac{1}{\sqrt{3 N}}\,, 
\end{equation}
since $|a| \ll 1$.
Thus the expected error on $\langle \cos\theta \rangle$ from just
statistics is 0.17 for Kamiokande II and 0.20 for IMB.  The range of
antineutrino energies contributing only negligibly increases the
error.  Thus the experimental results for both Kamiokande II and IMB
deviate by $+2$-$\sigma$ from the expectations.  (The disagreement
between the experimental results and the expectations is also
discussed in Refs.~\cite{Perkins,other87A}).  At the same time,
however, after correcting for the energy difference and the IMB
angular bias, the two means are in good agreement with each other.

It is generally assumed in the literature that the angular
distributions of Kamiokande II and IMB are consistent.  For example,
Ref.~\cite{LoSecco} claims an 81\% Kolmogorov-Smirnov probability that
the distributions are the same.  That test is primarily sensitive to
differences in means~\cite{NumRec}, and so this confirms the agreement
of the $\langle \cos\theta \rangle$ values above.  However, the
angular distribution is also characterized by its variance $\langle
\cos^2\theta \rangle - \langle \cos\theta \rangle^2$, with the
expectation being $\simeq 1/3$.  The error on the variance
is~\cite{Kendall}:
\begin{equation}
\delta\left(\langle \cos^2\theta \rangle - 
\langle \cos\theta \rangle^2\right)
\simeq \frac{1}{\sqrt{N}} \sqrt{\frac{1}{5} - \frac{1}{3^2}}
\simeq \frac{0.30}{\sqrt{N}}\,.
\end{equation}
The experimental result for the variance in Kamiokande II is 0.32,
with expected error 0.09, and hence in excellent agreement with
expectation.  The experimental result for IMB is 0.11, with expected
error 0.11, and hence a $-2$-$\sigma$ deviation from expectation and,
more importantly, from the Kamiokande II result.  Thus, contrary to
general belief, the Kamiokande II and IMB angular distributions,
characterized here by their first two moments, are {\it not
consistent} at the 2-$\sigma$ level.

It is possible that some of the observed forward events were due to
neutrino-electron scattering, though the expectation is only $\simeq
0.3$ events for Kamiokande II and $\simeq 0.1$ events in IMB (using
the same supernova parameters as in Ref.~\cite{SKpaper} and the
detector properties taken from Refs.~\cite{KamII,IMB}).  Most other
authors have also obtained an expectation of $\lesssim 1$ event per
detector.  Allowing $n$ neutrino-electron scattering events out of a
total of $N$ events will change the expectations for the mean $\langle
\cos\theta \rangle$ (increased by $n/N$) and the variance $\langle
\cos^2\theta \rangle - \langle \cos\theta \rangle^2$ (increased by
$2/3 (n/N) - (n/N)^2$).  Thus for $N \simeq 10$ and the possible $n =
1$, the means would be somewhat improved (now each a $+1.5$-$\sigma$
deviation).  The Kamiokande II variance would still be in agreement
with expectation, though the IMB variance would then be a
$-3$-$\sigma$ deviation.

As a general caution about the small-number statistics, one can
consider, for the purpose of illustration, the effect of assuming that
one backward event was missed.  That is, to each data set we add one
fake backward event.  For Kamiokande II, the effect on both the mean
and the variance is modest, but for IMB there is a large effect,
making the mean only a $+1$-$\sigma$ deviation and putting the
variance in agreement with theory.  Thus the statistical significance
can be very sensitive to small fluctuations, so that the number of
sigmas of deviation and the implicit confidence levels should be taken
with caution.

In conclusion, the Kamiokande II data seem to require a 2-$\sigma$
statistical fluctuation in the mean, and the IMB data separate
2-$\sigma$ fluctuations in the mean and the variance.  It is difficult
to explain the observed angular distributions, even taking into
account the corrections of Fig.~2 and (somewhat implausibly) allowing
$\simeq 1$ neutrino-electron scattering event in each detector.  Given
the inconsistency between the Kamiokande II and IMB angular
distributions, it is probably not legitimate to combine them, thus
weakening the argument for new supernova or particle physics that
could have affected the angular distributions, as invoked in
Refs.~\cite{LoSecco,vanderVelde,Danka}.  While we have not explained
the angular distributions, we have explicitly shown the perils of the
small-number statistics and an apparent additional problem with the
IMB results.

%%%%%%%%%%%%%%%%%%%%%%%%%%%%%%%%%%%%%%%%%%%%%%%%%%%%%%%%%%%%%%%%%%%%%%%%%%%%

\subsection{Supernova antineutrinos}

A strong $\bar{\nu}_e$ signal is expected in SuperKamiokande
\cite{SuperK} and other underground detectors from a future Galactic
supernova (for the expected count rates see, e.g.,
Refs.~\cite{SKpaper,SNOpaper}).  Is it possible to use the observed
angular distribution of the positron events to locate the direction of
the supernova~\cite{SNpointing}?  If the $M \rightarrow \infty$ limit
were appropriate (i.e., Eq.~(\ref{eq:avgcos0})), then the positrons
would be dominantly moving in the backward direction.  However, the
$1/M$ corrections calculated above are very important.  Folding in the
expected Fermi-Dirac distribution of the incoming $\bar{\nu}_e$, and
weighting $\cos\theta$ properly with the flux and cross section
calculated here, we arrive at $\langle \cos\theta \rangle \simeq$
0.015, 0.025 and 0.034 for temperatures $T$ = 4, 5, and 6 MeV.  Thus
the positrons from supernova $\bar{\nu}_e$, with most probable
temperature of about 5 MeV, will in fact be slightly forward and,
moreover the asymmetry coefficient will sensitively depend on the
antineutrino temperature, quite different from the naive expectation.
For locating the supernova, the best strategy seems to be to
concentrate on the positrons of the highest energy.  For $T = 5$ MeV,
about 25\% of the signal will be above 30 MeV, and should have a
noticeable forward asymmetry ($\langle \cos \theta \rangle$ = 0.056).
Observation of the angular distribution of the higher energy positrons
would constitute an important check of the supernova origin of the
signal, and would allow location of the supernova to about
$\delta(\cos\theta) \simeq 0.2$~\cite{SNpointing}.

If the supernova direction is known, then knowledge of the positron
angular distribution could be used to separate these events from other
reactions.  For example, if there are $\nu_\tau \rightarrow \nu_e$
oscillations, then the reaction $\nu_e + ^{16}{\rm O} \rightarrow e^-
+ ^{16}{\rm F}$ can become important~\cite{Haxton} (since the
$\nu_\tau$ temperature is higher than the $\nu_e$ temperature and this
reaction has a relatively high threshold).  The outgoing electrons are
somewhat backward.  Note that the neutron in $\bar{\nu}_e + p
\rightarrow e^+ + n$ is not detected, and electrons and positrons are
indistinguishable by their \v{C}erenkov radiation.  Therefore, the
search for events from $\nu_e + ^{16}{\rm O} \rightarrow e^- +
^{16}{\rm F}$ must be done statistically, by looking at the total
angular distribution and looking for a backward excess over what is
expected from $\bar{\nu}_e + p \rightarrow e^+ + n$ alone.  The
calculation in Ref.~\cite{Qian} used the naive positron asymmetry
$a^{(0)}$ and would have to be revised.  Since there would be fewer
backward events than they expected, the sensitivity would be improved.

In a heavy water detector like the Sudbury Neutrino Observatory
\cite{SNO}, the angular distributions of the outgoing leptons in the
reactions $\bar{\nu}_e + d \rightarrow e^+ + n + n$ and $\nu_e + d
\rightarrow e^- + p + p$ could also be used to locate the supernova
direction.  Because of the low numbers of events, however, even with a
naive asymmetry of $a^{(0)} = -1/3$, the pointing resolution is only
about $\delta(\cos\theta) \simeq 0.5$~\cite{SNpointing}.  Taking into
account the $1/M$ effects weakens the positron asymmetry, and would
degrade the pointing.  However, the corrected angular distributions
will still be quite important for separating reactions.

%%%%%%%%%%%%%%%%%%%%%%%%%%%%%%%%%%%%%%%%%%%%%%%%%%%%%%%%%%%%%%%%%%%%%%%%%%%%

\subsection{Search for solar antineutrinos}

Ref.~\cite{FMV} discusses the possibility of searching in
SuperKamiokande for $\bar{\nu}_e$ antineutrinos from the Sun,
presumably from $\nu_e \rightarrow \bar{\nu}_e$ oscillations, with
$E_\nu \sim 10$ MeV.  The authors proposed that these events could be
separated statistically by their angular distribution from the
isotropic detector background and the forward-peaked solar neutrino
events from neutrino-electron scattering.  However, from Fig.~2, the
angular asymmetry is substantially weaker than they assumed, and in
view of that, their derived limit would have to be modified.

%%%%%%%%%%%%%%%%%%%%%%%%%%%%%%%%%%%%%%%%%%%%%%%%%%%%%%%%%%%%%%%%%%%%%%%%%%%%

\subsection{LSND results}

Another important application of the positron angular distribution is
the search for neutrino oscillations by the LSND~\cite{LSND} and
KARMEN~\cite{KARMEN} collaborations.  The LSND collaboration reported
evidence for $\bar{\nu}_\mu \rightarrow \bar{\nu}_e$ oscillations
following $\mu^+$ decay at rest.  The evidence is based on the
observation of 22 $e^+$ + neutron events with positron energies
between 36 and 60 MeV, presumably originating from the $\bar{\nu}_e p$
interaction.  The directions of individual candidate positron events
have been measured with respect to the beam axis and the angular
distribution is given in Fig.~21 of Ref.~\cite{LSND}.  The measured
$\langle \cos\theta \rangle$ was found to be $0.20 \pm 0.13$.

From our Fig.~2, one would estimate that the expected value would be
only about 0.08.  However, there are important experimental
corrections particular to LSND which must be taken into account.  The
physics basis of most of them is the simple fact that, for a fixed
antineutrino energy $E_{\nu}$, the forward-going positrons have more
energy than the backward-going ones.  Thus, if there is a cut on
positron energy, say $E_{e^+} \ge 36$ MeV, then at the lowest allowed
antineutrino energies, only the forward positrons will survive the
cut, biasing $\langle \cos\theta \rangle$ to be larger.  (A similar
effect occurs due to the energy-dependent efficiency for positron
detection.)  These and related effects increase the expected value of
$\langle \cos\theta \rangle$ to 0.16, in good agreement with
observation~\cite{pcLouis}.  This suggests that LSND is indeed
observing $\bar{\nu}_e + p \rightarrow e^+ + n$ events, whatever the
origin of the $\bar{\nu}_e$. 

%%%%%%%%%%%%%%%%%%%%%%%%%%%%%%%%%%%%%%%%%%%%%%%%%%%%%%%%%%%%%%%%%%%%%%%%%%%%

\subsection{Relation to neutron beta decay parameters}

It is worth noting that the parameter $a(E_\nu)$ is the correlation
coefficient $a$ between the positron and antineutrino momenta in
neutron beta decay ~\cite{JTW}.  The parameter $a$ is difficult to
measure in neutron beta decay, since the antineutrino momentum can be
reconstructed only by measuring the proton recoil momentum.  The last
measurement~\cite{Stratowa}, more than 20 years ago, yielded $a =
-0.102 \pm 0.005$.  It is tempting to speculate that it might be
alternatively measured with $\bar{\nu}_e + p \rightarrow e^+ + n$ via
measurement of the positron energy and angle.

It must be pointed out, however, that even at modest $E_\nu$, weak
magnetism is an important correction, and thus a measurement of
$a(E_\nu)$ would probe a combination of $f$, $g$, and $f_2$, thus
providing instead a possible test of weak magnetism.  To pursue this
speculation further, consider an experiment in which $\langle
\cos\theta \rangle$ could be measured at a fixed $E_\nu$, and consider
how well the value of $(f + f_2)$ could be measured.  That is, an
experimental test of whether the value of $(f + f_2)$ extracted
matches the value (4.706) predicted by the Conserved Vector Current
hypothesis~\cite{Gell-Mann}.  The expectation for $\langle \cos\theta
\rangle$ is approximately given by Eq.~(\ref{eq:avgcos1a}).  From
statistics alone, $\delta(\langle \cos\theta \rangle) \simeq 1/\sqrt{3
N}$ for $N$ events, so that
\begin{equation}
\delta(f + f_2) =
\frac{\partial (f + f_2)}{\partial \langle \cos\theta \rangle}
\delta(\langle \cos\theta \rangle)
\simeq \frac{2}{E_\nu/M} \frac{1}{\sqrt{N}}\,.
\end{equation}
We assume the standard $V - A$ theory and neglect the (small) form
factor variation.  Some previous tests of weak magnetism were made by
measuring extremely small distortions in the beta spectra of the $A =
12$ and $A = 20$ systems (see Refs.~\cite{Telegdi,Holstein} for a
review).  These experiments reached a precision on $(f + f_2)$ of
about 10\%.  

As noted above, for a Galactic supernova at 10 kpc, observed in
SuperKamiokande, $\simeq 10^4$ $\bar{\nu}_e + p \rightarrow e^+ + n$
events are expected.  The time distribution of events is of course
irrelevant.  The spectrum $f(E_\nu)$ is also irrelevant insofar as it
will be determined from the data, since the measured $E_e$ and
$\cos\theta_e$ can be used to reconstruct $E_\nu$ for each event.  At
each value of $E_\nu$, $\langle \cos\theta \rangle$ and hence $(f +
f_2)$ could be measured.  While the typical energy expected is $\simeq
20$ MeV, about 2/3 of the events are at higher energies.  Thus one
might plausibly expect that a test of weak magnetism at the $\simeq
20\%$ level might be made (note also that the error varies linearly
with the assumed supernova distance).

%%%%%%%%%%%%%%%%%%%%%%%%%%%%%%%%%%%%%%%%%%%%%%%%%%%%%%%%%%%%%%%%%%%%%%%%%%%%
%%%%%%%%%%%%%%%%%%%%%%%%%%%%%%%%%%%%%%%%%%%%%%%%%%%%%%%%%%%%%%%%%%%%%%%%%%%%

\section{The neutron angular distribution and applications}

It is also of interest to consider the angular distribution of the
neutrons from $\bar{\nu}_e + p \rightarrow e^+ + n$, since the
neutrons are often detected as well.  In fact, the observation of the
neutron capture in a delayed coincidence with the positron is the
distinguishing signature of the antineutrino interaction with protons,
allowing suppression of backgrounds.

There is an angular correlation between the antineutrino direction and
the initial direction of the neutron.  Since in the laboratory system
the proton is at rest, momentum conservation requires that
\begin{equation}
\vec{p}_{\nu}  = \vec{p}_e + \vec{p}_n\,.
\end{equation}
Also,
\begin{equation}
|p_e| \le \sqrt{( E_{\nu} - \Delta )^2 - m_e^2 }  < E_{\nu}\,,
\end{equation}
so that the neutron must {\it always} be emitted in the forward
hemisphere.  In fact, the maximum angle $(\theta_n)_{max}$ between the
antineutrino and initial neutron directions is achieved when the
neutron and and positron momenta are perpendicular.  Neglecting
${\cal O}(1/M)$, this is
\begin{equation}
\cos(\theta_n)_{max} =
\frac{\sqrt{2 E_\nu \Delta - (\Delta^2 - m_e^2)}}{E_\nu}\,.
\end{equation}
At threshold, the neutron direction is purely forward, and at reactor
energies, still largely so.  In Fig.~4 we plot the quantity
$\cos(\theta_n)_{max}$ as well as the average $\langle \cos(\theta_n)
\rangle$, both evaluated numerically, where the latter was weighted
with the differential cross section.  At ${\cal O}(1/M)$ (see
Eq.~(\ref{eq:eel1})), the neutron kinetic energy is
\begin{equation}
T_n = \frac{E_\nu E_e^{(0)}}{M} \left(1 - v_e^{(0)}\cos\theta\right)
+ \frac{y^2}{M}\,.
\end{equation}

\begin{figure}[t]
\epsfxsize=3.25in \epsfbox{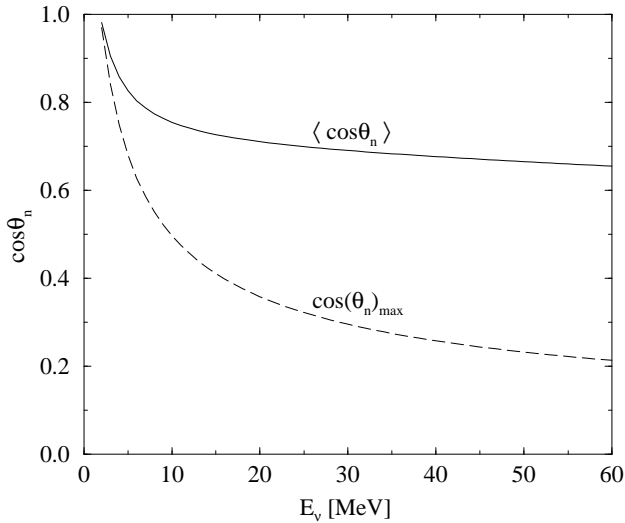}
\caption{The average neutron cosine and the cosine of the maximum
neutron angle versus antineutrino energy.  Note that
$\cos(\theta_n)_{max} \rightarrow 1$ both at threshold and as $E_\nu
\rightarrow \infty$.}
\end{figure}

It is often possible to localize, at least crudely, the points where
positron annihilated (essentially the point of creation for targets
with an appreciable density) and where the neutron was captured.  Even
though the neutron is captured only after many elastic scatterings,
its final position maintains some memory of its initial direction, as
we now show.  For a monoenergetic source of neutrons moving initially
along the $x$-axis, the distribution of final positions is Gaussian
distributed, with equal widths $\sigma_x$, $\sigma_y$, and $\sigma_z$.
The average final position $\langle x \rangle$ is displaced from the
origin, and $\langle y \rangle = \langle z \rangle = 0$.

We show the results of our Monte Carlo simulation in Fig.~5,
implemented by following the principles given by Fermi~\cite{Fermi}.
We assumed a liquid of $\left({\rm CH_2}\right)_n$, with or without Gd
doping (0.1\% by mass), and a density of 0.80 g/cm$^3$.  The results
in Fig.~5 can easily be rescaled to another density by multiplying by
$(0.80/\rho)$.  Neutrons are moderated by elastic scattering until
they reach thermal energy.  At thermal energy, elastic scattering
changes the neutron direction, but, on average, not its energy.  We
also implement capture on protons and Gd.  The calculated capture
times on undoped and doped scintillator are in good agreement with
expectation.

\begin{figure}[t]
\epsfxsize=3.25in \epsfbox{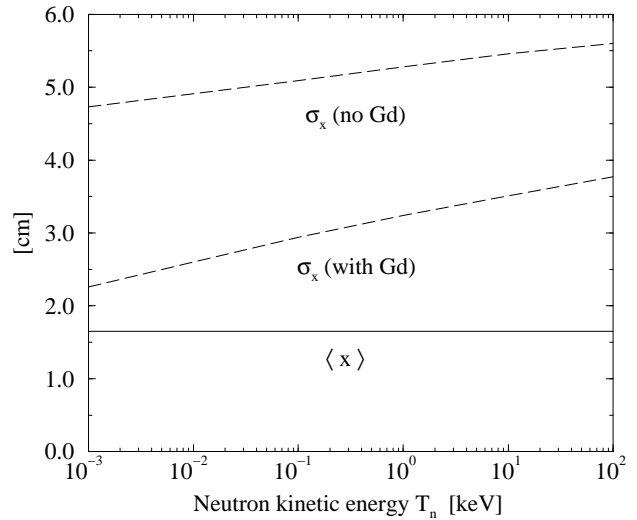}
\caption{The shift (solid line) $\langle x \rangle$ and width
(long-dashed line) $\sigma_x = \protect\sqrt{\langle x^2 \rangle -
\langle x \rangle^2}$ for monoenergetic neutrons (initial kinetic
energy $T_n$) emitted from the origin, moving initially along the
$x$-axis.  Note $\langle y \rangle = \langle z \rangle = 0$, and
$\sigma_x = \sigma_y = \sigma_z$.  We used a $\left(\protect{\rm
CH_2}\right)_n$ liquid of density 0.80 g/cm$^3$, with or without 0.1\%
Gd doping by mass.}
\end{figure}

The most significant input is the fact that the neutron elastic
scattering cross sections are almost independent of energy from about
10 keV down to about 0.1 eV~\cite{Ncross}.  For the first $\sim 5$
scatterings, the neutron maintains some sense of its initial
direction, with the angular distribution at a given scattering
depending only on the ratio of incoming and outgoing energies.  This
stage determines $\langle x \rangle \simeq 1.7$ cm, the average
displacement of the final point from the starting point.  All
subsequent scatters only enlarge the size of the neutron cloud.  The
width $\sigma_x$ does depend on the the initial neutron energy because
at higher energies more scatterings are required to moderate the
neutron to thermal energy.

Above 10 keV, the variation of the cross sections with energy should
be taken into account; doing so would increase $\langle x \rangle$ and
$\sigma_x$ somewhat for $T_n \simeq 100$ keV and more substantially
for higher energies.  Below about 0.1 eV, there is a chemical binding
effect depending on the moderating material that increases the cross
section; taking that into account would make $\sigma_x$ smaller.
Finally, from Fig.~4, the struck neutron is not exactly forward, as
assumed, but has $\cos\theta_n \simeq 0.9$ for reactor energies.
Taking that into account would reduce our shift to $\langle x \rangle
\simeq 1.5$ cm, in agreement with the results of
Ref.~\cite{Bemporad98}.

In fact, in the G\"{o}sgen reactor antineutrino
experiment~\cite{Gosgen} the neutron displacement was clearly
observed, at $\simeq 10 \sigma$ level~\cite{Zacek}.  This was possible
because the detector was composed of alternating walls of scintillator
and $^{3}{\rm He}$ neutron detectors.  For a given wall of
scintillator in which the reaction occurred, and the positron was
detected, more neutrons were observed in the $^{3}{\rm He}$ slab {\it
away} from the reactor than {\it towards} the reactor (in fact, the
ratio was about 2:1).  A similar effect was observed~\cite{Cussonneau}
in the Bugey 3 experiment~\cite{Bugey3}, also using a segmented
detector.

The neutron-positron separation is also being used~\cite{Bemporad98}
by the Chooz experiment~\cite{Chooz}, which is based on an unsegmented
detector.  The neutron position is only detected with a precision of
about 20 cm, but nevertheless a statistically significant displacement
of positron and neutron detection positions along the antineutrino
direction is seen.

Given a reliable calculation of the neutron transport in the detector,
and hence the expected neutron distributions, this technique would
allow a direct determination of the detector background from the
measured asymmetry.  Such an analysis is being pursued for the Palo
Verde reactor experiment, and a forward-backward asymmetry between
different cells is seen in the current data~\cite{pcBoehm}.

%%%%%%%%%%%%%%%%%%%%%%%%%%%%%%%%%%%%%%%%%%%%%%%%%%%%%%%%%%%%%%%%%%%%%%%%%%%%
%%%%%%%%%%%%%%%%%%%%%%%%%%%%%%%%%%%%%%%%%%%%%%%%%%%%%%%%%%%%%%%%%%%%%%%%%%%%

\section{Conclusions}

We have given an expression for the differential cross section in the
positron $\cos\theta$, valid to order $1/M$.  Recoil and weak
magnetism corrections have a large energy-dependent effect on the
positron angular distribution, changing it from slightly backward at
low energies to isotropic at about 15 MeV and slightly forward at
higher energies.  

Our main result for the total cross section, valid to order $1/M$, is
obtained by integrating Eq.~(\ref{eq:dsig1}).  At low energies, this
agrees with the well-known $M \rightarrow \infty$ result.  At high
energies, where the threshold can be neglected, this is in good
agreement with Eq.~(3.18) of Ref.~\cite{LS}, which contains all orders
in $1/M$ but assumes $\Delta = 0$.  Using our result, we have
determined the largest $\Delta$-dependent terms missing in the results
of Ref.~\cite{LS}.  The most accurate formula for the total cross
section at all energies is obtained by using the result of
Ref.~\cite{LS} with our modifications.  The modified formula is
essentially identical to our main result for $E_\nu \lesssim 30$ MeV
and is in good agreement with it (and the unmodified result of
Ref.~\cite{LS}) at higher energies.

The positron angular distribution is well-described by $\langle
\cos\theta \rangle$.  Our main result, Eq.~(\ref{eq:avgcos1}), valid
to order $1/M$, is an excellent approximation over the entire energy
range considered (and in fact up to about $E_\nu \simeq 150$ MeV).

A number of experimental applications are discussed in which the
correct angular distribution is necessary for separating $\bar{\nu}_e
+ p \rightarrow e^+ + n$ events from other reactions and from detector
backgrounds.

The neutron angular distribution is initially strongly forward.  The
random walk of the neutrons as they are moderated acts to erase this.
However, the centroid of the final distribution of neutron capture
positions is shifted in the direction of the initial motion.  We
discuss how this can be exploited experimentally.

%%%%%%%%%%%%%%%%%%%%%%%%%%%%%%%%%%%%%%%%%%%%%%%%%%%%%%%%%%%%%%%%%%%%%%%%%%%%
%%%%%%%%%%%%%%%%%%%%%%%%%%%%%%%%%%%%%%%%%%%%%%%%%%%%%%%%%%%%%%%%%%%%%%%%%%%%

\newpage

\section*{ACKNOWLEDGMENTS}

This work was supported in part by the U.S. Department of Energy under
Grant No. DE-FG03-88ER-40397.  J.F.B. was supported by a Sherman
Fairchild fellowship from Caltech.  We also thank the Aspen Center for
Physics, where part of this work was done.  We thank Gerry Garvey and
Bill Louis for discussions of the angular distribution of the
candidate $\bar{\nu}_e$ events observed in LSND; Brian Fujikawa and
Gerry Garvey for discussions of weak magnetism; Carlo Bemporad, Felix
Boehm, Yves Declais, and Andreas Piepke for discussions of the
positron-neutron separation in the detection of reactor antineutrinos;
and Kuniharu Kubodera for providing and explaining his numerical
tables of the charged-current neutrino-deuteron differential cross
sections.

%%%%%%%%%%%%%%%%%%%%%%%%%%%%%%%%%%%%%%%%%%%%%%%%%%%%%%%%%%%%%%%%%%%%%%%%%%%%
%%%%%%%%%%%%%%%%%%%%%%%%%%%%%%%%%%%%%%%%%%%%%%%%%%%%%%%%%%%%%%%%%%%%%%%%%%%%

\appendix

\section{Kinematic relations}

In the center of momentum frame, the threshold is defined by the
positron and neutron being produced at rest, so
\begin{equation}
E_\nu^{thr.} = 
\frac{(M_n + m_e)^2 - M_p^2}{2 (M_n + m_e)} = 1.803 {\rm\ MeV}\,.
\end{equation}
In the laboratory frame (where the proton is at rest), the threshold
is
\begin{equation}
E_\nu^{thr.} = 
\frac{(M_n + m_e)^2 - M_p^2}{2 M_p} = 1.806 {\rm\ MeV}\,.
\end{equation}

Labeling the 4-momenta as $\bar{\nu}_e(p_\nu) + p(p_p) \rightarrow
e^+(p_e) + n(p_n)$, we define the Mandelstam variables as
\begin{eqnarray}
s & = & (p_\nu + p_p)^2 = M_p^2 + 2 M_p E_\nu \\
t & = & (p_\nu - p_e)^2 = M_n^2 - M_p^2 - 2 M_p (E_\nu - E_e) \\
u & = & (p_\nu - p_n)^2 = M_p^2 + m_e^2 - 2 M_p E_e \,,
\end{eqnarray}
evaluated in the laboratory frame, where we can also write $t = q^2 =
m_e^2 - 2 E_\nu E_e (1 - v_e \cos\theta)$.

The differential cross section in $t$ can be written as
\begin{equation}
\frac{d \sigma}{d t} = \frac{G_F^2 \cos^2\theta_C}{\pi}
\frac{|{\cal M}|^2}{(s - M_p^2)^2}\; (1 + \Delta^R_{inner})\,,
\end{equation}
where $|{\cal M}|^2$ is the amplitude squared (averaged over initial
spins, summed over final spins).  This can be written as the
differential cross section in the positron $\cos\theta$ in the
laboratory by using the Jacobian
\begin{equation}
\frac{d t}{d \cos\theta} = 2 E_\nu p_e^{(1)}
\left[1 - \frac{E_\nu}{M}
\left(1 - \frac{1}{v_e^{(0)}} \cos\theta\right)
+ {\cal O}\left(\frac{1}{M^2}\right)\right]\,.
\end{equation}
The differential cross section in $E_e$ can be obtained by using $d
q^2/d E_e = 2 M_p$.

%%%%%%%%%%%%%%%%%%%%%%%%%%%%%%%%%%%%%%%%%%%%%%%%%%%%%%%%%%%%%%%%%%%%%%%%%%%%
%%%%%%%%%%%%%%%%%%%%%%%%%%%%%%%%%%%%%%%%%%%%%%%%%%%%%%%%%%%%%%%%%%%%%%%%%%%%

%%%%%%%%%%%%%%%%%%%%%%%%%%%%%%%%%%%%%%%%%%%%%%%%%%%%%%%%%%%%%%%%%%%%%%%%%%%%
%%%%%%%%%%%%%%%%%%%%%%%%%%%%%%%%%%%%%%%%%%%%%%%%%%%%%%%%%%%%%%%%%%%%%%%%%%%%

\end{document}